\documentclass[twocolumn,showpacs,preprintnumbers,amsmath,amssymb]{revtex4}
\usepackage{color}

\newcommand{\ket}[1]{\mid\!#1\,\rangle}

\newcommand{\sss}[1]{\scriptscriptstyle #1}

\begin{document}

\title{Canonical $SO(2,4)$-invariant quantization in conformally flat spaces}

\author{
Sofiane Faci
}
\email{sofiane@cbpf.br}
\affiliation{
Centro Brasileiro de Pesquisas Fisicas (CBPF-ICRA),
Rua Dr. Xavier Sigaud, 150, Urca, CEP 22290-180, Rio de Janeiro, RJ, Brasil
}

\date{\today}% It is always \today, today,

\begin{abstract}
We show how to quantize $SO(2,d)$-invariant fields in $d>2$ dimensional conformally flat spaces (CFS). The Weyl equivalence between CFSs is exploited to perform the quantization process in Minkowski space then transport the entire $SO(2,d)$-invariant structure to curved CFSs. 
We make use of the canonical quantization scheme and a special careful is made to specify a scalar product, technically related to a Cauchy surface.
The latter is chosen to be common to all globally hyperbolic CFSs in order to relate the different associated Hilbert spaces. The quantum fields are constructed and the two-point functions are given in terms of their minkowskian counterparts. It appears that an $SO(2,d)$-invariant quantum field does not locally distinguish between two different CFSs. 
\end{abstract}


\maketitle

\section{Introduction}
Conformal invariance is a fundamental ingredient for modern theoretical physics and covers several areas, see for instance \cite{Kastrup} and references therein.
There is a rich literature dealing with Weyl invariance \cite{ORaifeartaigh:2000, Scholz:2011fk, Castro:2009zza, Jackiw:2005su}
and also with $SO(2,d)$-invariant field theories (CFT) in  $d>2$ spaces \cite{Weinberg:2010,  Maldacena:2011jn, Antoniadis:2011, Nikolov:2007, Todorov:2012xx, Costa:2011}.
The case of $d=2$ is particular since the conformal group becomes infinite \cite{cft-2d} and, exploring the Polyakov formalism, was at the origin of String Theory \cite{CFT-Strings}.
On the other hand, the set of conformally flat spaces (CFS) is of great importance. Indeed it includes Minkowski, FLRW and de Sitter spaces. According to the standard cosmological model, our Universe has been in a de Sitter phase  during inflation and has afterwords gone through an expanding - currently accelerating - FLRW space, it might come back to a de Sitter shape in the far future \cite{Carroll:1997, Mukhanov:2005}. Moreover, CFSs represent the first step toward general curved manifolds. Thus, well understand the quantization process in these spaces is a capital step towards quantum field theory in general curved spaces and quantum gravity.

Weyl and the restricted conformal group $SO(2,d)$ transformations are different but subtilely related. Indeed, a $d>2$ dimensional  classical Weyl invariant field theory restricted to live in a CFS yields an $SO(2,d)$-invariant field theory \cite{pconf7}. %This generalizes Zumino's work related to Minkowski space  \cite{Zumino}.
The demonstration of this result relies on the $SO(2,4)$-invariant structure transport of a classical field theory from Minkowski space to a CFS. 
The present paper extends this approach to include quantum fields. The main goal is to construct standard $SO(2,d)$-invariant quantum field theories in arbitrary CFSs. 

For this purpose, we use the canonical quantization scheme which goes by three steps:
i) Construct a Hilbert space
ii) Define a unitary representation
iii) Find a causal reproducing kernel.

A Hilbert space is a complete set of normalized modes, relative to a scalar product which is technically related to a Cauchy surface. The key point of the present work is that it is possible to choose the same Cauchy surface for all CFSs \cite{Fulling}. This allows to have a one-to-one correspondance between all Hilbert spaces in the different CFSs and consequently to simply relate the Wightman functions.
The idea is to explore the Weyl equivalence between CFSs to perform the quantization process in Minkowski space then transport the entire $SO(2,d)$-invariant structure to curved CFSs. 
This is possible since the $SO(2,4)$ is the smallest group containing as subgroups all isometry groups associated to the CFSs \cite{Keane:1999}. Also, an $SO(2,4)$-invariant field theory in a CFS is also invariant under the associated isometry group, which means a well defined theory.

%\paragraph{\textbf{The Weyl Bridge.}}
To develop, let $\cal M$ be a real $d>2$ dimensional differential manifold. 
Minkowski space reads $(\mathcal{M}, \eta_{\mu\nu})$ where $ \eta_{\mu\nu}=(+,--...)$.
An arbitrary CFS  $(\mathcal{M}, \bar g_{\mu\nu})$ is locally related to Minkowski space by a Weyl rescaling 
\begin{equation}\label{weyl}
\bar{g}_{\mu\nu} = K^2\,  \eta_{\mu\nu},
\end{equation}
where the Weyl factor $K$ is a real, non vanishing $C^\infty$ function.
The considered spaces are globally hyperbolic, otherwise Cauchy problem is not well defined. In such spaces, the present formalism can be applied to bounded subspaces where a Cauchy surface can be defined.
Our methodology is based on two maps: the map $W$ to transport the fields and the map $H$ to transport the differential operators acting on these fields. That is, for a giving conformal field $F$ and an arbitrary operator $\cal O$, defined in Minkowski space, the two maps read
\begin{eqnarray}
W : F  & \rightarrow & \bar{F} = W F \label{W}
\\
H : \mathcal{O}  & \rightarrow & \bar{\mathcal{O}} = W \mathcal{O} W^{-1} \label{H},
\end{eqnarray}
where $W$ denotes the map and its matrix representation. 
The map (\ref{W}) is usually defined as $\bar{F}=K^s\, F$ (where $s$ is called the conformal weight) but can be extended to include a tensorial part. 
Say, for a tensor field with components $F_{\sss I}$, the general form of the matrix $W$ reads
%\begin{equation}
%\begin{split}
$\bar{F}_{\sss I} %& = W_{\sss I}^{\sss J} F_{\sss J}
%\\& 
= K^s \left( \delta_{\sss I}^{\sss J} + \Upsilon_{\sss I}^{\sss J} \right) F_{\sss J} ,
$ %\end{split}
%\end{equation}
where $\Upsilon_{\sss I}^{\sss J}$ is a non-diagonal matrix which depends on the function $K$ and its derivatives. Explicit examples of such matrices can be found in \cite{pconf3, pconf6}.
As a consequence, the field $\bar F$ and the operator $\bar {\cal O}$ are defined in the CFS 
$(\mathcal{M}, \bar g_{\mu\nu})$.

%\newpage

\section{Canonical quantization in Minkowski space}
\subsection{A classical theory}
Let us start by considering a classical free field $F$ defined in Minkowski space and obeying to the $SO(2,4)$-invariant equation 
\begin{equation}\label{equation-F}
\mathcal{E}F = 0,
\end{equation}
where $\mathcal{E}$ is a given differential operator. 
Furthermore we assume that this equation does not need boundary conditions - other than initial data on some Cauchy surface - to be solved.
The $SO(2,4)$-invariance means
\begin{equation}\label{invariance-equation}
\forall e \in SO(2,4), \qquad  {\cal E} \, U_{e} F(x) = w_{e}(x) \,  {\cal E} \, F(x),
\end{equation}
where 
\begin{equation}\label{rep}
 U_e F(x) = M_e(x) F(e^{-1}.x),
\end{equation}
where $w_{e}(x)$ and $M_e(x)$ are matrices.
This representation can also be expressed in an infinitesimal form by the commutation relations between the group generators $X_{e}\in so(2,4)$ and the field $F$
\begin{equation}\label{action-F}
[X_{e}, F] = X_{e} F + \Sigma_{e} F,
\end{equation}
where $X_{e}F$ denotes the scalar action and $\Sigma_{e}F$ the tensorial action.
The infinitesimal form of (\ref{invariance-equation}) reads
\begin{equation}\label{invariance-equation-infinitesimal}
\forall X_{e} \in so(2,4), \qquad [ {\cal E}, X_{e} ] = \xi_{e} \, {\cal E},
\end{equation}
where $\xi_{e}$ are real functions.
Since $SO(2,4)$ is a Lie group, the finite transformations are obtained from the infinitesimal ones by the exponential application,
\begin{equation}\label{exp}
 U_e F(x) = e^{\lambda_e^{i} X_e^{i}} F(x).
\end{equation}

\subsection{A quantum theory}
The $SO(2,4)$-covariant quantization is achieved by constructing a Hilbert space - on which $SO(2,4)$ acts unitarily - and an invariant Wightman function.

A Hilbert space is a complet set of normalized modes ${\cal H}=\{F_k\}$. These are solutions of Eq. (\ref{equation-F}) and normalized, 
\begin{equation}\label{ps-general}
 \langle F_{k},\ F_{k'}\rangle=\delta_{kk'},
\end{equation}
%where $\delta_{kk'}$ is the Cronecker symbol.
according to a scalar product $\langle, \rangle$. To ensure unitarily the scalar product should be $SO(2,4)$-invariant,  that is
\begin{equation}\label{ps-general}
\forall e \in SO(2,4), \qquad  \langle U_e F_{k},\ U_e F_{k'}\rangle=\langle F_{k},\ F_{k'}\rangle.
\end{equation}

%When the scalar product is indefinite, the space  $\cal H$ is then of a Krein type \cite{Garidi:2004}.

The Wightman two-point function reads
\begin{equation}\label{wi}
\begin{split}
{\cal W}(x,x') 
%& = \langle 0 \mid  \hat F(x) \hat F(x') \ket{0}  
%\\
& = \sum_k   F_{k}^*(x)\ F_{k}(x'),
\end{split}
\end{equation}
%when $x$ and $x'$ are causally related, otherwise ${\cal W}=0$. 
and provides a causal and covariant reproducing kernel of $\cal H$:
\begin{itemize}
\item A reproducing kernel.
\begin{equation}\label{noyau}
\forall F\in {\cal H}, \quad \langle {\cal W}(x, \cdot),F\rangle = F(x).
\end{equation}
\item 
Causality.
 \begin{equation}\label{}
 {\cal W}(x,x')={\cal W}(x',x)
 \end{equation}
  as soon as $x$ and $x'$ are causally separated.
 \item
 $\cal W$ is  $SO(2,4)$-invariant. This comes from the relations (\ref{ps-general}), (\ref{wi}) and (\ref{noyau}).
\end{itemize}

Afterwards the quantum field $\hat F$ is constructed as
\begin{equation}\label{F-quantique}
\hat{F}(x)= \hat a ({\cal W}(x,.)) + \hat a^\dag({\cal W}(x,.)) ,
\end{equation}
where $\hat a$ and $\hat a^\dag$ are respectively anti-linear and linear operators - actually, they are operator-valued tempered distributions - acting on a Fock space.

As a consequence, using (\ref{wi}), the above field can be expanded as
\begin{equation}\label{F-quantique}
\hat{F}(x)=\sum_k \, F_{k}(x) \hat a_k+ F_{k}^*(x) \hat a^\dag_k,
\end{equation}
where $\hat a_k = \hat a(F_{k})$ and $\hat a^\dag_k = a^\dag(F_{k})$ are the standard annihilation and creation operators of the modes $F_k$ and obeying the canonical commutation relations (ccr) 
\begin{equation}\label{algebre-g}
\begin{split}
& [\hat a_{k} , \hat a_{k'}] =[ \hat a_{k}^\dag , \hat a_{k'}^\dag] = 0, %\quad  [\hat a_{k} , \hat a_{k'}^\dag] = \epsilon_{k} \delta_{kk'}.
\\
&  [\hat a_{k} , \hat a_{k'}^\dag] = \epsilon_{k} \delta_{kk'},
\end{split}
\end{equation} 
where $\epsilon_{k}=\pm1$. In the case $\epsilon_{k}$ does not have the same sign for all modes  $F_{k}$, the space  $\cal K$ is then of a Krein type.
Note that the appearance of the non-trivial algebraic structure indicates the entrance into the quantum world.

The one particle sector of the Fock space, denoted $\underline{\cal H}$, contains the quantum states 
\begin{equation}
\ket{F_{k}} %= \hat a_{k}^\dag \ket{0} 
= \hat a^\dag(F_{k}) \ket{0},
\end{equation}
where the conformal vacuum state $\ket{0}$ verifies
\begin{equation}\label{vacuum}
\forall \hat a_{k}, \quad \hat a_{k} \ket{0} = 0,
\end{equation}
and is the unique invariant state of $\underline{\cal H}$,
\begin{equation}\label{vacuum}
\forall e\in SO(2,4), \quad  \underline{U}_e \ket{0} = \ket{0}.
\end{equation}
So the $SO(2,4)$-invariance permits to pick up a preferred vacuum state, which corresponds to a preferred basis of ${\cal H}=\{F_k\}$. This solves one of the most persistent difficulties in QFT in curved spaces.

The Wightman function (\ref{wi}) can thus be written under the usual form
\begin{equation}
{\cal W}(x,x')  = \langle 0 \mid  \hat F(x) \, \hat F(x') \ket{0}  .
\end{equation}

The resulting quantum field has the following properties.
\begin{itemize}
\item
It verifies the field equation
\begin{equation}
\forall \, F_1, F_2 \in {\cal H}, \qquad \langle F_{1} \mid {\cal E}\hat{F} \mid F_{2} \rangle = 0.
\end{equation}
For more details see the appendix of \cite{pconf3}.

\item
Is causal. If $x$ and $x'$ are causally disjoint then
\begin{equation}\label{causality}
[\hat F(x), \hat F(x')] = 0.
\end{equation}

\item
Is $SO(2,4)$-covariant
\begin{equation}
\forall \, e \in SO(2,4), \qquad \underline{U}^{-1}_e \, F(x) \, \underline{U}_e = F(e.x).
\end{equation}

\end{itemize}

\section{Going to a CFS}
\subsection{Transporting the classical theory}
The two maps $W$ and $H$ allow to transport the equation (\ref{equation-F}) to the CFS. The new equation reads
\begin{equation}\label{equation-F-bar}
\begin{array}{ccl}
\bar{\mathcal{E}} \bar F & =& H {\cal E}\ W F
\\
 & =  &   W \mathcal{E} W^{-1} \, W F 
\\
  & =  &   W (\mathcal{E} F)
\\
 & = & 0.
\end{array}
\end{equation} 
The CFS generators of $so(2,4)$ result from applying the map $H$ on $\{X_{e}\}$:
\begin{equation}
X_{e} \to \bar{X}_{e}= H X_{e} = W X_{e} W^{-1}.
\end{equation}
They act on the field $\bar F$ as
\begin{equation}\label{action-transport}
\begin{split}
[\bar X_{e}, \bar F] & = \bar X_{e} \bar F + \bar \Sigma_{e} \bar F 
%\\ & 
%= H X_{e} \  \bar F  + H \Sigma_{e}  \ \bar F
%\\ & 
%= W X_{e} W^{-1} \  \bar F  +  \Sigma_{e}  \ \bar F
\\ & 
= W X_{e} \ F  +  \Sigma_{e}  W \ F
\\ & 
=  W\, [ X, F ],
\end{split}
\end{equation}
where $\bar \Sigma_{e} = \Sigma_{e}$ because the spinorial action $\Sigma_{e}$ does not contain derivatives.
Yet the finite transformations are obtained by the exponential application,
\begin{equation}\label{rep-fini}
\begin{split}
\bar U_e \bar F(x) 
& = e^{\lambda_e^{i} \bar X_e^{i}} \bar F(x),
\\
& = W(x) \, e^{\lambda_e^{i} X_e^{i}} F(x),
\\
& = W(x) \, U_e F(x),
\\&
= W(x) \, M_e(x) F(e^{-1}.x),
\\&
= W(x) \, M_e(x) \, [W(e^{-1}.x)]^{-1}\, \bar F(e^{-1}.x),
\\&
=\bar M_e(x)\, \bar F(e^{-1}.x).
\end{split}
\end{equation}
The Minkowskian and CFS  representations are related through
 $\bar M_e(x)= W(x) \, M_e(x) \, [W(e^{-1}.x)]^{-1}$.

The invariance of Eq. (\ref{equation-F-bar}) follows from the third line of (\ref{rep-fini}) and yields, $\forall e \in SO(2,4),$ 
\begin{equation}\label{}
\begin{split}
   \bar {\cal E} \bar U_{e} \bar F(x) 
      & =  W  {\cal E} \, U_{e} F(x) ,
\\
   & =  W  \, w_{e}(x) \,   {\cal E} \, F(x),
\\
   & =  w_{e}(x) \,  \bar {\cal E} \, \bar F(x).
\end{split}
\end{equation}
The infinitesimal invariance is easier to implement
\begin{equation}\label{inv-eq-bar}
\begin{array}{ccl}
\forall X_{e} \in so(2,4), \qquad
[\bar{\mathcal{E}}, \bar{X}_{e}]   
%  & = & W [\mathcal{E}, X_{e}] W^{-1}
%  \\
  & = &  W \, \zeta_{e} \mathcal{E} \, W^{-1}
  \\
  & = & \zeta_{e} \, \bar{\mathcal{E}}.
\end{array}
\end{equation}
%which is the counterpart of (\ref{invariance-equation-infinitesimal}).
Note that the identity
$\bar{\zeta}_{e}=\zeta_{e},$
or equivalently $\bar  w_{e}=  w_{e},$ means that the resulting CFS theory keeps the same $SO(2,4)$-invariant structure of the minkowskian theory.

\subsection{Transporting the quantum theory}
Let us turn to the quantum field. 
The modes $\{F_{k}\}$, solutions of (\ref{equation-F}), are transported using the map $W$ to get the modes
\begin{equation}\label{ }
\bar F_{k} = W F_{k},
\end{equation}
solutions of the CFS equation (\ref{equation-F-bar}). 
The set of these new modes $\{\bar F_{k}\}$ forms a basis for a new Hilbert space denoted $\bar {\cal H}$. This space is equipped with a new scalar product $\langle\langle , \rangle\rangle$, defined in such a way to ensure the normalization of the modes $\bar F_{k}$ in the same way as for the modes $F_{k}$:
\begin{equation}\label{ }
\langle \langle  \bar F_{k} , \bar F_{k'} \rangle\rangle = \langle F_{k}, F_{k'} \rangle = \epsilon_{k}\, \delta_{kk'}.
\end{equation}
This is possible since we are free to choose the same Cauchy surface to define both minkowskian and CFS scalar products \cite{Fulling}.
Moreover the scalar product $\langle\langle, \rangle\rangle$ is $SO(2,4)$-invariant,
\begin{equation}\label{ps-general-cfs}
 \langle\langle U_e \bar F_{k},\ U_e \bar F_{k'}\rangle\rangle=\langle\langle \bar F_{k},\ \bar F_{k'}\rangle\rangle,
\end{equation}
which means that here again $U$ acts unitarily on $\bar {\cal H}$.

The Wightman two-point function reads
\begin{equation}\label{transport-wightman}
\begin{split}
\bar {\cal W}(x,x') 
& = \sum_k  \, \bar F_{k}^*(x)\  \bar F_{k}(x'),
\\&
 =  W(x) W(x') \ {\cal W}(x,x'),
\end{split}
\end{equation}
%when $x$ and $x'$ are causally related, otherwise ${\cal W}=0$. Thus,
which automatically provides a causal and covariant reproducing kernel of $\bar {\cal H}$:
\begin{itemize}
\item A reproducing kernel,
\begin{equation}\label{noyau-bar}
\forall \bar F\in \bar {\cal H}, \quad \langle\langle \bar {\cal W}(x, \cdot),\bar F\rangle\rangle = \bar F(x).
\end{equation}
\item 
Causality. It comes from that of $\cal W$ and the fact that a Weyl rescaling preserve the space causal structure. That is
 \begin{equation}\label{}
\bar  {\cal W}(x,x')= \bar {\cal W}(x',x)
 \end{equation}
  as soon as $x$ and $x'$ are causally separated.
 \item
Finally, $\bar {\cal W}$ is  $SO(2,4)$-invariant. This comes from the unitarity condition (\ref{ps-general-cfs}).
\end{itemize}

The CFS quantum field is then constructed with the new Wightman function,
\begin{equation}\label{F-quantique}
\hat{\bar{F}}(x)= \hat a (\bar {\cal W}(x,.)) + \hat a^\dag(\bar {\cal W}(x,.)),
\end{equation}
which yields
\begin{equation}
\bar {\cal W}(x,x')  = \langle \bar 0 \mid  \hat {\bar F}(x) \, \hat{\bar F}(x') \ket{\bar 0}.
\end{equation}

The vacuum state $\ket{\bar 0}$ is the conformal one. It is the unique invariant state of $\underline{\bar {\cal H}}$,
\begin{equation}\label{vacuum-cfs}
\forall e\in SO(2,4), \quad  \underline{\bar {U}}_e \ket{\bar 0} = \ket{\bar 0}.
\end{equation}

As a consequence, using (\ref{wi}), the above field can be expanded as
\begin{equation}\label{F-quantique-tr}
\hat{\bar F}(x)=\sum_k \, \bar F_{k}(x) \hat a_k + \bar F_{k}^*(x) \hat a^\dag_k,
\end{equation}
where the annihilators and creators are formally identical to the Minkowskian ones, verifying the algebra (\ref{algebre-g}), but acting on the modes $\{\bar F_k\}$: $\hat a_k = \hat a(\bar F_{k})$, $\hat a^\dag_k = a^\dag(\bar F_{k})$ and thus verifying $\hat a_{k} \ket{\bar 0} = 0$.
The one particle sector of the CFS Fock space $\underline{\bar {\cal H}}$ contains the quantum states 
\begin{equation}
\ket{\bar F_{k}} %= \hat a_{k}^\dag \ket{0} 
= \hat a^\dag(\bar F_{k}) \ket{\bar 0}.
\end{equation}

The resulting quantum field
\begin{itemize}
\item
Verifies the field equation
\begin{equation}
\forall \bar F_{1}, \bar F_{2} \in \bar{\cal H}, \quad
\langle \bar F_{1} \mid {\bar{\cal E}}\hat{\bar F} \mid \bar F_{2} \rangle = 0.
\end{equation}

\item
Causal. If $x$ and $x'$ are causally disjoint then
\begin{equation}\label{causality}
[\hat{\bar F}(x), \hat {\bar F}(x')] = 0.
\end{equation}

\item
$SO(2,4)$-covariant
\begin{equation}
\forall \, e \in SO(2,4), \qquad \bar{\underline{U}}^{-1}_e \, \bar F(x) \, \bar {\underline{U}}_e = \bar F(e.x).
\end{equation}

\end{itemize}

Note that the quantum fields $\hat F$ and $\hat {\bar F}$, nor the quantum states $\ket{F_k}\in\underline{{\cal H}}$ and $\ket{\bar F_k}\in \bar{\underline{{\cal H}}}$, are not explicitly related. Though this does not prevent the Wightman functions to be related through
\begin{equation}
\bar {\cal W}(x,x')  =  W(x) W(x') \ {\cal W}(x,x').
\end{equation}
This is the crux of the present paper. This is well known for the scalar field but not in the general case that was demonstrated here.

\section{Concluding remarks}
We have succeeded in formally transport the whole $SO(2,4)$-invariant structure of a minkowskian quantum field theory to obtain a new $SO(2,4)$-invariant theory defined in an arbitrary conformally flat space. 
Two key points were explored. i) It is possible to choose the same Cauchy surface in all CFSs in order to define both minkowskian and curved CFS scalar products.
ii) The $SO(2,4)$ group is the smallest group containing as subgroups all isometry groups associated to the CFSs.

The $SO(2,4)$-invariant quantum structure in two CFSs relies on the same quantum operators acting on different but one-to-one related Hilbert spaces.
 It appears that an $SO(2,4)$-invariant quantum field does not locally distinguish between two different CFSs.
This is particularly true for high-frequency modes which only "interact" with the close neighborhood.

Note that the Weyl rescaling (\ref{weyl}) and thus the maps (\ref{W}) and  (\ref{H}) are local and depend on a coordinate system, which does not, in general, cover the whole spaces. Nevertheless the $SO(2,4)$ is a Lie group and the global action can be obtained from the infinitesimal one using the exponential application. Moreover, in case of need, several coordinate systems can be used to cover the whole spaces (examples are given in \cite{Higuchi:2008, pconf4}).

Note also that only free fields were considered in this work, also no interaction-like conformal anomalies could appear \cite{Drummond:1979uq}. 
Nonetheless free-like conformal anomalies can appear in bounded curved spaces - like de Sitter space - \cite{Wald:1978, Tsoupros:fk}. These are important and problematic global effects that come from the gravitational interaction feedback. The present work focused on local properties and did not treat conformal anomalies.
% This task can be connected to the works \cite{Kraus:1992ru, Kraus:1992th} and provides future investigations.

%\tableofcontents

\baselineskip=10pt
\bibliography{Biblio}

\begin{thebibliography}{25}
\expandafter\ifx\csname natexlab\endcsname\relax\def\natexlab#1{#1}\fi
\expandafter\ifx\csname bibnamefont\endcsname\relax
  \def\bibnamefont#1{#1}\fi
\expandafter\ifx\csname bibfnamefont\endcsname\relax
  \def\bibfnamefont#1{#1}\fi
\expandafter\ifx\csname citenamefont\endcsname\relax
  \def\citenamefont#1{#1}\fi
\expandafter\ifx\csname url\endcsname\relax
  \def\url#1{\texttt{#1}}\fi
\expandafter\ifx\csname urlprefix\endcsname\relax\def\urlprefix{URL }\fi
\providecommand{\bibinfo}[2]{#2}
\providecommand{\eprint}[2][]{\url{#2}}

\bibitem[{\citenamefont{Kastrup}(2008)}]{Kastrup}
\bibinfo{author}{\bibfnamefont{H.}~\bibnamefont{Kastrup}},
  \bibinfo{journal}{Annalen Phys.} \textbf{\bibinfo{volume}{17}},
  \bibinfo{pages}{631} (\bibinfo{year}{2008}), \eprint{0808.2730}.

\bibitem[{\citenamefont{O'Raifeartaigh and
  Straumann}(2000)}]{ORaifeartaigh:2000}
\bibinfo{author}{\bibfnamefont{L.}~\bibnamefont{O'Raifeartaigh}}
  \bibnamefont{and}
  \bibinfo{author}{\bibfnamefont{N.}~\bibnamefont{Straumann}},
  \bibinfo{journal}{Rev.Mod.Phys.} \textbf{\bibinfo{volume}{72}},
  \bibinfo{pages}{1} (\bibinfo{year}{2000}).

\bibitem[{\citenamefont{Scholz}(2011)}]{Scholz:2011fk}
\bibinfo{author}{\bibfnamefont{E.}~\bibnamefont{Scholz}}
  (\bibinfo{year}{2011}), \eprint{arxiv: 1102.3478v2}.

\bibitem[{\citenamefont{Castro}(2009)}]{Castro:2009zza}
\bibinfo{author}{\bibfnamefont{C.}~\bibnamefont{Castro}},
  \bibinfo{journal}{J.Global Sci.Tech.} \textbf{\bibinfo{volume}{1}},
  \bibinfo{pages}{1} (\bibinfo{year}{2009}).

\bibitem[{\citenamefont{Jackiw}(2006)}]{Jackiw:2005su}
\bibinfo{author}{\bibfnamefont{R.}~\bibnamefont{Jackiw}},
  \bibinfo{journal}{Theor.Math.Phys.} \textbf{\bibinfo{volume}{148}},
  \bibinfo{pages}{941} (\bibinfo{year}{2006}), \eprint{hep-th/0511065}.

\bibitem[{\citenamefont{Weinberg}(2010)}]{Weinberg:2010}
\bibinfo{author}{\bibfnamefont{S.}~\bibnamefont{Weinberg}},
  \bibinfo{journal}{Phys. Rev.} \textbf{\bibinfo{volume}{D82}},
  \bibinfo{pages}{045031} (\bibinfo{year}{2010}), \eprint{1006.3480}.

\bibitem[{\citenamefont{Maldacena and Zhiboedov}(2011)}]{Maldacena:2011jn}
\bibinfo{author}{\bibfnamefont{J.}~\bibnamefont{Maldacena}} \bibnamefont{and}
  \bibinfo{author}{\bibfnamefont{A.}~\bibnamefont{Zhiboedov}}
  (\bibinfo{year}{2011}), \eprint{1112.1016}.

\bibitem[{\citenamefont{Antoniadis~Ignatios and Emil}(2011)}]{Antoniadis:2011}
\bibinfo{author}{\bibfnamefont{M.~P.~O.} \bibnamefont{Antoniadis~Ignatios}}
  \bibnamefont{and} \bibinfo{author}{\bibfnamefont{M.}~\bibnamefont{Emil}}
  (\bibinfo{year}{2011}), \eprint{1103.4164}.

\bibitem[{\citenamefont{Nikolov et~al.}(2007)\citenamefont{Nikolov, Rehren, and
  Todorov}}]{Nikolov:2007}
\bibinfo{author}{\bibfnamefont{N.~M.} \bibnamefont{Nikolov}},
  \bibinfo{author}{\bibfnamefont{K.-H.} \bibnamefont{Rehren}},
  \bibnamefont{and} \bibinfo{author}{\bibfnamefont{I.}~\bibnamefont{Todorov}},
  \bibinfo{journal}{http://arxiv.org/abs/0711.0628}  (\bibinfo{year}{2007}),
  \eprint{0711.0628}.

\bibitem[{\citenamefont{Todorov}(2012)}]{Todorov:2012xx}
\bibinfo{author}{\bibfnamefont{I.}~\bibnamefont{Todorov}}
  (\bibinfo{year}{2012}), \eprint{1207.3661}.

\bibitem[{\citenamefont{Costa et~al.}(2011)\citenamefont{Costa, Penedones,
  Poland, and Rychkov}}]{Costa:2011}
\bibinfo{author}{\bibfnamefont{M.~S.} \bibnamefont{Costa}},
  \bibinfo{author}{\bibfnamefont{J.}~\bibnamefont{Penedones}},
  \bibinfo{author}{\bibfnamefont{D.}~\bibnamefont{Poland}}, \bibnamefont{and}
  \bibinfo{author}{\bibfnamefont{S.}~\bibnamefont{Rychkov}},
  \bibinfo{journal}{JHEP} \textbf{\bibinfo{volume}{1111}}, \bibinfo{pages}{071}
  (\bibinfo{year}{2011}), \eprint{1107.3554}.

\bibitem[{\citenamefont{Belavin et~al.}(1984)\citenamefont{Belavin, Polyakov,
  and Zamolodchikov}}]{cft-2d}
\bibinfo{author}{\bibfnamefont{A.}~\bibnamefont{Belavin}},
  \bibinfo{author}{\bibfnamefont{A.}~\bibnamefont{Polyakov}}, \bibnamefont{and}
  \bibinfo{author}{\bibfnamefont{A.}~\bibnamefont{Zamolodchikov}},
  \bibinfo{journal}{Nuclear Physics B} \textbf{\bibinfo{volume}{241}},
  \bibinfo{pages}{333 } (\bibinfo{year}{1984}).

\bibitem[{\citenamefont{Blumenhagen and Plauschinn}(2009)}]{CFT-Strings}
\bibinfo{author}{\bibfnamefont{R.}~\bibnamefont{Blumenhagen}} \bibnamefont{and}
  \bibinfo{author}{\bibfnamefont{E.}~\bibnamefont{Plauschinn}},
  \emph{\bibinfo{title}{Introduction to conformal field theory: With
  application to string theory}}, vol. \bibinfo{volume}{779} of
  \emph{\bibinfo{series}{Lecture notes in Physics}}
  (\bibinfo{publisher}{Springer}, \bibinfo{year}{2009}).

\bibitem[{\citenamefont{Carroll}(1997)}]{Carroll:1997}
\bibinfo{author}{\bibfnamefont{S.~M.} \bibnamefont{Carroll}},
  \emph{\bibinfo{title}{{Lecture notes on general relativity}}}
  (\bibinfo{year}{1997}), \eprint{gr-qc/9712019}.

\bibitem[{\citenamefont{Mukhanov}(2005)}]{Mukhanov:2005}
\bibinfo{author}{\bibfnamefont{V.}~\bibnamefont{Mukhanov}}
  (\bibinfo{year}{2005}).

\bibitem[{\citenamefont{Faci}(2013)}]{pconf7}
\bibinfo{author}{\bibfnamefont{S.}~\bibnamefont{Faci}}, \bibinfo{journal}{EPL}
  (\bibinfo{year}{2013}).

\bibitem[{\citenamefont{Fulling}(1989)}]{Fulling}
\bibinfo{author}{\bibfnamefont{S.}~\bibnamefont{Fulling}},
  \emph{\bibinfo{title}{{ASPECTS OF QUANTUM FIELD THEORY IN CURVED
  SPACE-TIME}}}, vol.~\bibinfo{volume}{17} (\bibinfo{year}{1989}).

\bibitem[{\citenamefont{Keane and Barrett}(2000)}]{Keane:1999}
\bibinfo{author}{\bibfnamefont{A.~J.} \bibnamefont{Keane}} \bibnamefont{and}
  \bibinfo{author}{\bibfnamefont{R.~K.} \bibnamefont{Barrett}},
  \bibinfo{journal}{Class.Quant.Grav.} \textbf{\bibinfo{volume}{17}},
  \bibinfo{pages}{201} (\bibinfo{year}{2000}), \eprint{gr-qc/9907002}.

\bibitem[{\citenamefont{Faci et~al.}(2009)\citenamefont{Faci, Huguet, Queva,
  and Renaud}}]{pconf3}
\bibinfo{author}{\bibfnamefont{S.}~\bibnamefont{Faci}},
  \bibinfo{author}{\bibfnamefont{E.}~\bibnamefont{Huguet}},
  \bibinfo{author}{\bibfnamefont{J.}~\bibnamefont{Queva}}, \bibnamefont{and}
  \bibinfo{author}{\bibfnamefont{J.}~\bibnamefont{Renaud}},
  \bibinfo{journal}{Phys. Rev. D} \textbf{\bibinfo{volume}{80}},
  \bibinfo{pages}{124005} (\bibinfo{year}{2009}), \eprint{0910.1279}.

\bibitem[{\citenamefont{Faci}(2012)}]{pconf6}
\bibinfo{author}{\bibfnamefont{S.}~\bibnamefont{Faci}},
  \bibinfo{journal}{arXiv:1205.3804}  (\bibinfo{year}{2012}).

\bibitem[{\citenamefont{Higuchi and Lee}(2008)}]{Higuchi:2008}
\bibinfo{author}{\bibfnamefont{A.}~\bibnamefont{Higuchi}} \bibnamefont{and}
  \bibinfo{author}{\bibfnamefont{Y.~C.} \bibnamefont{Lee}},
  \bibinfo{journal}{Phys.Rev.} \textbf{\bibinfo{volume}{D78}},
  \bibinfo{pages}{084031} (\bibinfo{year}{2008}), \eprint{0808.0642}.

\bibitem[{\citenamefont{Faci et~al.}(2011)\citenamefont{Faci, Huguet, and
  Renaud}}]{pconf4}
\bibinfo{author}{\bibfnamefont{S.}~\bibnamefont{Faci}},
  \bibinfo{author}{\bibfnamefont{E.}~\bibnamefont{Huguet}}, \bibnamefont{and}
  \bibinfo{author}{\bibfnamefont{J.}~\bibnamefont{Renaud}},
  \bibinfo{journal}{Phys.Rev.} \textbf{\bibinfo{volume}{D84}},
  \bibinfo{pages}{124050} (\bibinfo{year}{2011}), \eprint{arxiv:1110.1177}.

\bibitem[{\citenamefont{Drummond and Shore}(1979)}]{Drummond:1979uq}
\bibinfo{author}{\bibfnamefont{I.}~\bibnamefont{Drummond}} \bibnamefont{and}
  \bibinfo{author}{\bibfnamefont{D.~M.} \bibnamefont{Shore}},
  \bibinfo{journal}{Physical Review D} \textbf{\bibinfo{volume}{19}},
  \bibinfo{pages}{1134} (\bibinfo{year}{1979}).

\bibitem[{\citenamefont{Wald}(1978)}]{Wald:1978}
\bibinfo{author}{\bibfnamefont{R.~M.} \bibnamefont{Wald}},
  \bibinfo{journal}{Phys.Rev.} \textbf{\bibinfo{volume}{D17}},
  \bibinfo{pages}{1477} (\bibinfo{year}{1978}).

\bibitem[{\citenamefont{Tsoupros}()}]{Tsoupros:fk}
\bibinfo{author}{\bibfnamefont{G.}~\bibnamefont{Tsoupros}},
  \bibinfo{journal}{arXiv:0409163}  (????), \eprint{hep-th/0409163}.

\end{thebibliography}

\end{document}